\documentclass[aps,prb,showpacs,superscriptaddress,twocolumn,floatfix]{revtex4-1}
\usepackage{epsfig}
\usepackage{microtype}
\usepackage{epstopdf}
\usepackage{amsmath, amsthm, amssymb}
\usepackage{hyperref}
\usepackage{color}

\def \d {\mathrm{d}}

\def \imp {{\rm imp}}
\def\l{\Lambda}
\def\a{\alpha}
\def\b{\beta}

\def\nn{\nonumber\\}
\def\eps{\varepsilon}
\newcommand{\be}{\begin{equation}}
\newcommand{\ee}{\end{equation}}
\newcommand{\bea}{\begin{eqnarray}}
\newcommand{\eea}{\end{eqnarray}}

\def\fr#1{(\ref{#1})}
\begin{document}
\title{Real-time dynamics in the one-dimensional Hubbard model}
\author{Luis Seabra}
\affiliation{Department of Physics, Technion - Israel Institute of
  Technology, Haifa 32000, Israel}
\author{Fabian H.L. Essler}
\affiliation{The Rudolf Peierls Centre for Theoretical Physics, Oxford
  University, Oxford OX1 3NP, UK} 
\author{Frank Pollmann}
\affiliation{
Max-Planck-Institut f\"ur Physik komplexer Systeme, N\"othnitzer
Stra\ss e 38, 01187 Dresden, Germany}
\author{Imke Schneider}
\affiliation{The Rudolf Peierls Centre for Theoretical Physics, Oxford University, Oxford OX1 3NP, United Kingdom}
\author{Thomas Veness}
\affiliation{The Rudolf Peierls Centre for Theoretical Physics, Oxford
  University, Oxford OX1 3NP, UK} 

\begin{abstract}
We consider single-particle properties in the one-dimensional
repulsive Hubbard model at commensurate fillings in the metallic
phase. We determine the real-time evolution of the retarded Green's
function by matrix-product state methods. We find that at sufficiently
late times the numerical results are in good agreement with
predictions of nonlinear Luttinger liquid theory. We argue that
combining the two methods provides a way of determining the
single-particle spectral function with very high frequency
resolution.
\end{abstract}


\maketitle

\section{Introduction}

The spectral and dynamical properties of one-dimensional  fermionic systems have attracted a significant amount of interest in recent years due to advances in both low-dimensional materials~\cite{Baeriswyl2004} and cold atom systems~\cite{Bloch2012}.
The repulsive Hubbard model constitutes a key paradigm for studying strong correlation
effects in these systems \cite{book}. Its
Hamiltonian is
\bea
H=&-&J\sum_{j,\sigma}c^\dagger_{j,\sigma}c_{j+1,\sigma}+
c^\dagger_{j+1,\sigma}c_{j,\sigma}
+U\sum_j n_{j,\uparrow}\ n_{j,\downarrow}\nn
&-&\mu\sum_j n_j,
\label{HHubb}
\eea
where $n_{j,\sigma}=c^\dagger_{j,\sigma}c_{j,\sigma}$ and $n_j=
n_{j,\uparrow}+n_{j,\downarrow}$.
In one dimension, it is exactly solvable via the Bethe
Ansatz\cite{book}, with many results about the model available in the
literature. 
However, the computation of dynamical correlation functions
analytically and directly from the Bethe Ansatz remains a difficult
task. An example of particular interest is the single-particle
spectral function
\bea
A(\omega,k)&=&-\frac{1}{\pi}\ {\rm Im}\ G_{\rm ret}(\omega,k),\nn
G_{\rm ret}(\omega,k)&=&-i\int_0^\infty dt\ e^{i\omega t}\sum_l e^{-ikla_0}\nn
&&\quad\times\ \langle \psi_0|\{c_{j+l,\sigma}(t),\ c^\dagger_{j,\sigma}\}|\psi_0\rangle,
\label{specfun}
\eea
where $|\psi_0\rangle$ is the ground state.
The spectral function is accessible through angle-resolved photoemission
experiments. Such measurements on the quasi-1D organic conductor
TTF-TCNQ have been interpreted in terms of $A(\omega,k)$ of the 1D
Hubbard model \cite{TTF,jeckel}. 

One approach to calculating properties of the spectral function 
is via Luttinger liquid
theory\cite{boso,thierry}. Certain low-energy aspects  of the Hubbard
model in zero magnetic field are described by the bosonized
\mbox{Hamiltonian~\mbox{$H = H_c + H_s$}} 
\bea
H=\sum_{\a=c,s}\frac{v_\a}{2\pi}\int dx\left[
\frac{1}{K_\a}\Big(\frac{\partial \Phi_\a}{\partial x}\Big)^2
+K_\a\Big(\frac{\partial \Theta_\a}{\partial x}\Big)^2\right]
+\ldots
\nn
\label{HFT}
\eea
Here the dots indicate the presence of \emph{irrelevant} operators,
which means that their respective
coupling constants approach zero under the renormalization group flow.
Neglecting the effects of the irrelevant operators in \fr{HFT} makes
it possible to calculate $A(\omega,k)$ at low energies
\cite{Lutt,boso,book,thierry}. The spectral 
function is found to exhibit singularities following the dispersions
of the collective spin (``spinon'') and charge (``holon'')
excitations. A problem with this Luttinger liquid approach to
dynamical correlation functions is that the latter inherently involve
a finite energy scale [e.g. the frequency $\omega$ in \fr{specfun}],
while the coupling constants of irrelevant perturbations \fr{HFT}
will, at best, be small but finite at the scale $\omega$. 

In a number of works
\cite{work1,Roz06,work2,BAW1,work3,work4,work5,IG08,work6,BAW2,ABW,work7,work8,Affleck2,pereira12,Roz14,austen},
it was demonstrated, for the case of spinless fermions, that
taking the irrelevant operators perturbing the Luttinger liquid
Hamiltonian into account generally leads to significant changes in
the singularities characterizing the dominant features seen in
response functions. 
Exact expressions for these singular
features in response functions were
obtained\cite{work1,Roz06,work2,BAW1,work3,work4,work5,IG08,work6,work7,work8,Affleck2},  using a mapping to a Luttinger liquid coupled to a
mobile impurity\cite{balents}.
Crucially, unlike in unperturbed Luttinger liquid theory, the
exponents characterizing these singularities are generally
\emph{momentum dependent}. The generalization to spinful fermions, and
the Hubbard model in particular, was considered in several recent
works\cite{PS,SIG,FHLE,Pereira,ts}. The case of the Mott-insulating phase
 beyond the field theory regime\cite{FTMI} was treated in
Ref.~[\onlinecite{Pereira}], and a rather complete understanding of
the dynamical structure factor was obtained.

%
\begin{figure*}[ht]
\includegraphics[width=16cm]{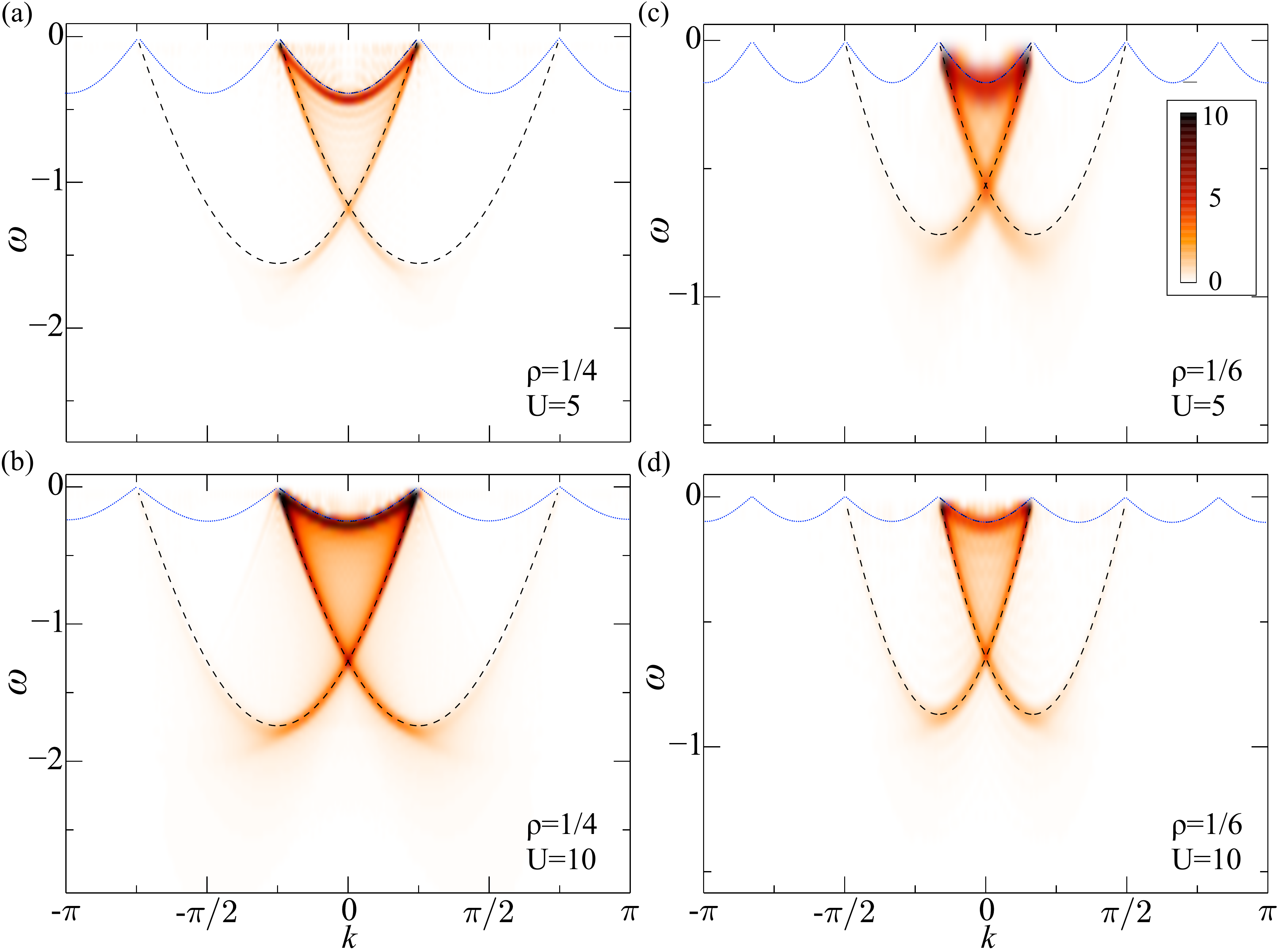}
\caption{(Color online)   Spectral function $A(\omega,k)$ of the
Hubbard model calculated with TEBD at filling $\rho=1/4$  for (a)
$U=5$ and (b)  $U=10$ and at filling $\rho=1/6$ for (a) $U=5$ and
(b) $U=10$. The black dashed lines are the singularity thresholds
obtained from the Bethe Ansatz solution. In all cases the spectral
function is non-zero everywhere below the absolute threshold indicated
by the thin blue lines, although the intensity is mostly very small.
\label{fig:Akw-n14}
}
\end{figure*}
%

For the metallic phase of the Hubbard model
(less than one electron per site), a mobile impurity model
 was formulated to describe the singular features in the single-particle spectral
function\cite{SIG}.
Combined with
the Bethe Ansatz solution, this allowed the calculation of exact expressions for the
power-law singularities occurring at the ``absolute threshold''\cite{FHLE}.
These results were found to be in accord with previous analytic work
based on completely different
assumptions and methods~\cite{carmelo}.
 On the other hand, a comparison of the
analytic predictions\cite{FHLE} with available dynamical density matrix renormalization group
(dDMRG)\cite{white1992,Jeckelmann2002} results on the
quarter-filled Hubbard model \cite{jeckel} was found to be rather
unconvincing.
 A serious complication is that the dDMRG method used
in Ref.~[\onlinecite{jeckel}] requires a finite imaginary part of the
frequency, while the window in frequency space over which the
calculated power law applies in general will be narrow. This
makes extracting power-law singularities from dDMRG results in the
momentum/frequency domain inherently difficult.

In this work, we employ a numerical approach to check the validity
of the results for threshold singularities in response functions
obtained by mobile impurity models. Using the time-evolving block
decimation method (TEBD), formally equivalent to time-dependent
density matrix renormalization group (tDMRG),
cf. Refs.~[\onlinecite{Vidal2004,White2008,Schollwock2011,Pereira}],
we compute the retarded Green's function
in the momentum/time domain for a variety of fillings and interaction
strengths.
As we will see, these results are well described by a fit
to a power-law decay expression whose frequencies and exponents are
fixed by combining Bethe Ansatz results with an appropriate mobile
impurity model. 

The paper is organized as follows. In section~\ref{sec:numerical-akw}
we present the single-particle spectrum through the spectral function
$A(\omega,k)$ calculated using TEBD.   
In section~\ref{sec:excitations} we identify its main features in
terms of the exact excitation spectrum known from Bethe
Ansatz, and present results for edge exponents
obtained from appropriate mobile impurity models.
Implications of these findings for real-time dynamics are summarized
in section~\ref{sec:impurity}.
Section~\ref{section:comparison} is devoted to a comparison of these
results to the numerically calculated Green's function in the momentum/time domain.
We summarize our results in section~\ref{sec:conclusions}.

\section{Numerical study of the spectral function}  \label{sec:numerical-akw}

We first use the DMRG algorithm to find the ground state $|\psi_0\rangle$ of the Hamiltonian (\ref{HHubb}) in an MPS representation. The TEBD algorithm is then used to obtain its dynamical properties
(details of the  numerical method used can be found in Refs~[\onlinecite{white1992,Vidal2004,Seabra2013,Sticlet2014,Kjaell2013}]). 
In particular, we calculate the retarded  single-particle Green's function $G_{\rm ret}(\omega,k)$=$G^{\rm -}(\omega,k)+G^{\rm +}(\omega,k)$ from the Fourier transform for positive times of the dynamical two-point correlation functions
\begin{align}
G^{\rm -}(t,j)=-i\langle \psi_0| c^{\dagger}_ {j}(t) c^{\phantom \dagger}_{j_0}(0) |\psi_0\rangle,
\label{eq:GFs}
\\
G^{\rm +}(t,j)=-i\langle  \psi_0|c^{\phantom \dagger}_{j}(t) c^{\dagger}_{ j_0}(0) |\psi_0\rangle.
\label{eq:GFs-2}
\end{align}
Each Green's function requires a separate simulation, performed by 
time-evolving an excited MPS, where an operator $c_{j_0}$ or $c_{j_0}^\dagger$ (the spin index is suppressed for clarity) has been applied at $t=0$ to to the ground state MPS, at the central site of a finite chain $j_0=L/2$, see e.g. \mbox{Refs.~[\onlinecite{Affleck2,Pereira,Schollwock2011}]}.
We use open chains of length $L=300$, and let the bond dimensions grow with time such that the truncation error is at most $10^{-5}$ per step. 
A fourth-order Suzuki-Trotter decomposition with time steps of $\sim0.05$ (in units of inverse hopping)  is employed.
The maximum time is chosen such that the wave-front of the ``light-cone" of correlations does not reach the edges of the system, which introduces a cutoff at small frequencies.
Before performing a Fourier transform of the data, the sampled time is extended using linear prediction~\cite{White2008}, improving the energy resolution.
In order to set the Fermi level at $\epsilon_F=0$, the chemical potential $\mu$ in~(\ref{HHubb}) is adjusted so that $E_0(N-1)=E_0(N)$ where $E_0(N)$ is the ground state energy with $N$ electrons.	

Throughout this paper we concentrate on the hole part of the Green's
function (\ref{eq:GFs}), and consequently the presented spectral
function only has support for negative frequencies $\omega<0$. This
spectral function is the quantity experimentally relevant to
photoemission spectra. 
The spectral function for  
 filling factors $\rho=1/4$ and $\rho=1/6$ with interaction strengths $U$=$5$ and $U$=$10$ is shown in Fig.~\ref{fig:Akw-n14}.
The main characteristics for $|k|<k_F$ can be clearly observed, as previously reported in e.g.~Ref.~[\onlinecite{jeckel}].
%
An injected fermion separates into (at least) one ``spinon'', a
gapless  spin-1/2 collective excitation with no charge, and one
``(anti-)holon'', a gapped, charged $\pm e$ collective excitation with
no spin. 
The sharp line of the several $A(\omega,k)$ in Fig.~\ref{fig:Akw-n14} near $\omega=0$ is the dispersion of the spinon excitation, while the lines below correspond to the holon excitations.
Along these lines, $A(\omega,k)$ behaves as a power-law singularity.
For a given total momentum there exist holon-spinon states in a range of frequencies.
%
These excitations are thoroughly described  in the next section from the point of view of Bethe Ansatz (dashed lines in  Fig.~\ref{fig:Akw-n14}).
In the remainder of this paper we will focus in analysing the threshold singularities at the sharp edges for $|k|<k_F$.
For  $|k|>k_F$ there is a strong decay in the spectral weight and further features in the spectral function are difficult to observe for these parameters.
%

\section{Excitations in the Hubbard model and dominant features in the
  spectral function}  \label{sec:excitations}
A detailed discussion of the excitation spectrum of the Hubbard model
is given in Ref.~[\onlinecite{book}] (see also
Refs.~[\onlinecite{excitations}]).
The particular
excitations relevant to the description of the single-particle Green's
function have been constructed in detail in Ref.~[\onlinecite{FHLE}],
and we now summarize the relevant results given there. We consider the
case of zero magnetic field and $N$ electrons on an $L$-site periodic
chain. The Fermi momentum is then
\be
k_F=\frac{\pi N}{2L}.
\ee
The dominant features in the (hole) spectral function at momentum
$|q|<k_F$ arise from the
{\sl holon-spinon} excitation carrying charge $+e$ and spin $1/2$. Its
energy and momentum are
\bea
E_{\rm hs}(k^h,\Lambda^h) &=& -\varepsilon_c(k^h) - \varepsilon_s(\Lambda^h)\ ,\nn
P_{\rm hs}(k^h,\Lambda^h) &=& -p_c(k^h) - p_s(\Lambda^h)\pm 2k_F\ ,
\label{EPhs}
\eea
where $|k^h|<Q$ and $-\infty<\Lambda^h<\infty$ parametrize the
excitation. The functions $\varepsilon_{c,s}$ and $p_{c,s}$ are
obtained from the solutions of coupled linear integral equations and
are given in (19) and (28) of Ref.~[\onlinecite{FHLE}] respectively.
The origin of the $\pm 2k_F$ contribution is discussed in Chapter
7.7.1. of Ref.~[\onlinecite{book}].
Because of parity invariance the Green's function is symmetric in
momentum and we therefore restrict ourselves to the momentum range
\be
0\leq P_{\rm hs}\leq k_F.
\ee

\subsection{Absolute Threshold: spinon edge}
%
%
The absolute threshold for $0<P_{\rm hs}<k_F$ was analyzed in detail
in Ref.~[\onlinecite{FHLE}]. It is obtained by choosing the plus sign
in \fr{EPhs}, fixing $k_h=Q$, and then varying $\Lambda^h$ in the range
\be
-\infty<\Lambda^h\leq 0.
\ee
At energies just above this threshold, the spectral function exhibits
a power-law singularity\cite{SIG,FHLE} (as a function of frequency for
fixed momentum)
\be
A\big(\omega,P_{\rm hs}(Q,\Lambda^h)\big)
\propto \left(\omega-E_{\rm hs}\big(Q,\Lambda^h)\big)\right)^{-\mu^s_{0,-}}\ ,
\ee
where the exponent $\mu^s_{0,-}$ is given in (129) of Ref.~[\onlinecite{FHLE}].

\subsection{Holon edge}
%
The other dominant features in the spectral function arise in the
vicinity of the holon edge, obtained by choosing the minus sign in
\fr{EPhs}, setting $\Lambda^h=-\infty$ and varying $k^h$ in the range
\be
-Q\leq k^h<0.
\ee
The range of the corresponding momentum is \mbox{$-k_F\leq P_{\rm
  hs}<k_F$}. We note that by virtue of parity invariance this
particular branch is sufficient for describing both high-energy
features in the spectral function $A(\omega,0<q<k_F)$. We now assume that,
as a consequence of the integrability of the Hubbard model, a threshold
singularity occurs just above the holon edge (as a function of frequency for
fixed momentum)
\be
A\big(\omega,P_{\rm hs}(k^h,-\infty)\big)
\propto \left(\omega-E_{\rm hs}\big(k^h,-\infty)\big)\right)^{-\mu^c_{0,-}}\ .
\ee
The assumption that in integrable models threshold singularities occur
even at thresholds of excitations that sit on top of continua, into
which they are kinematically allowed to decay, appears reasonable: for
massive integrable quantum field theories this has been seen to be the
case\cite{SGM}. In non-integrable models one does not expect
singular behaviour\cite{SIG}.
%
\begin{figure}[t!]
\begin{center}
\epsfxsize=0.48\textwidth
\epsfbox{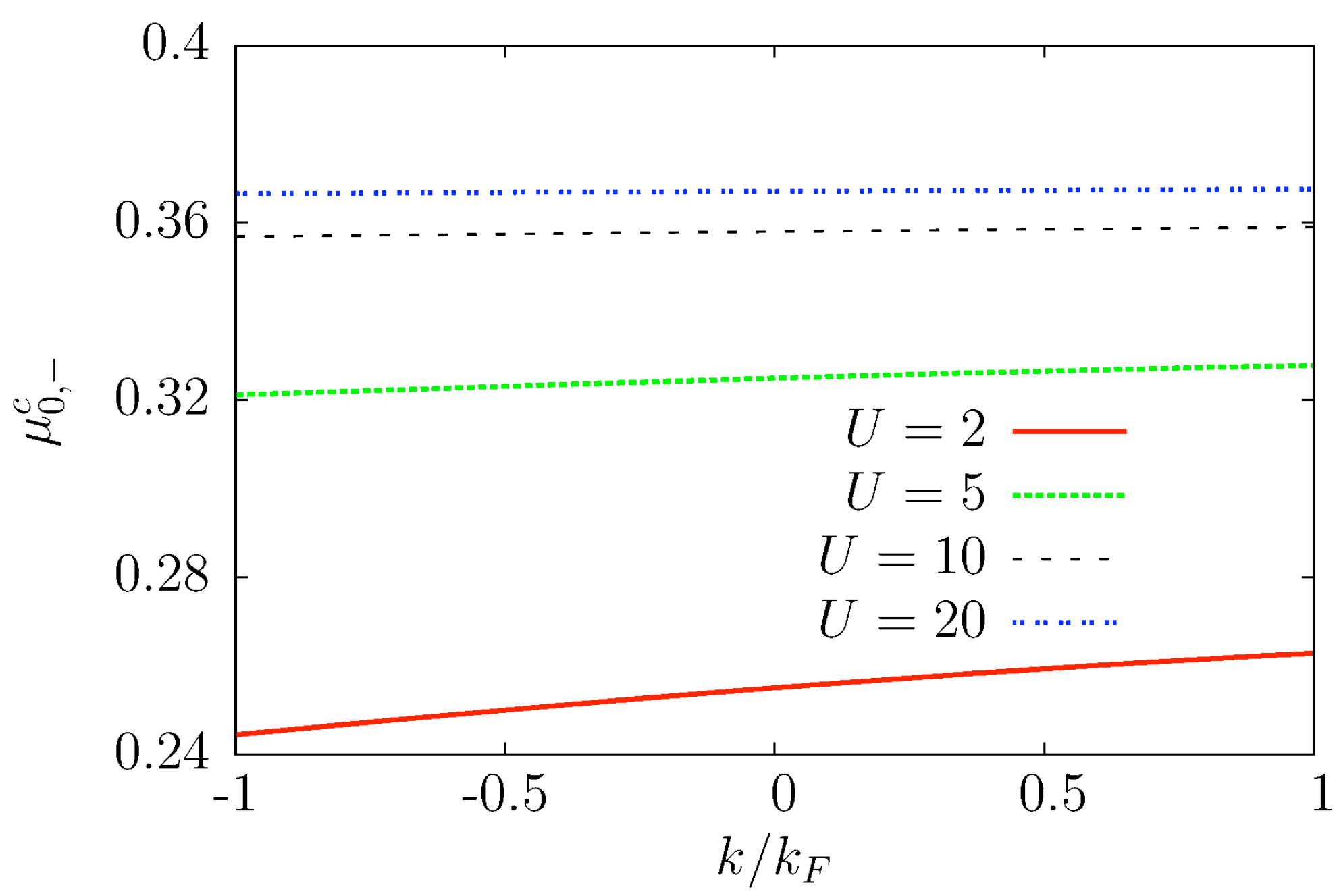}
\end{center}
\caption{(Color online)  
 Exponent $\mu_{0,-}^c$ for $\rho=1/4$ as a function of momentum as derived from the mobile impurity model approach for various fillings and values of $U$.}
\label{fig:rhop25} 
\end{figure}
%
%
\begin{figure}[ht]
\begin{center}
\epsfxsize=0.48\textwidth
\epsfbox{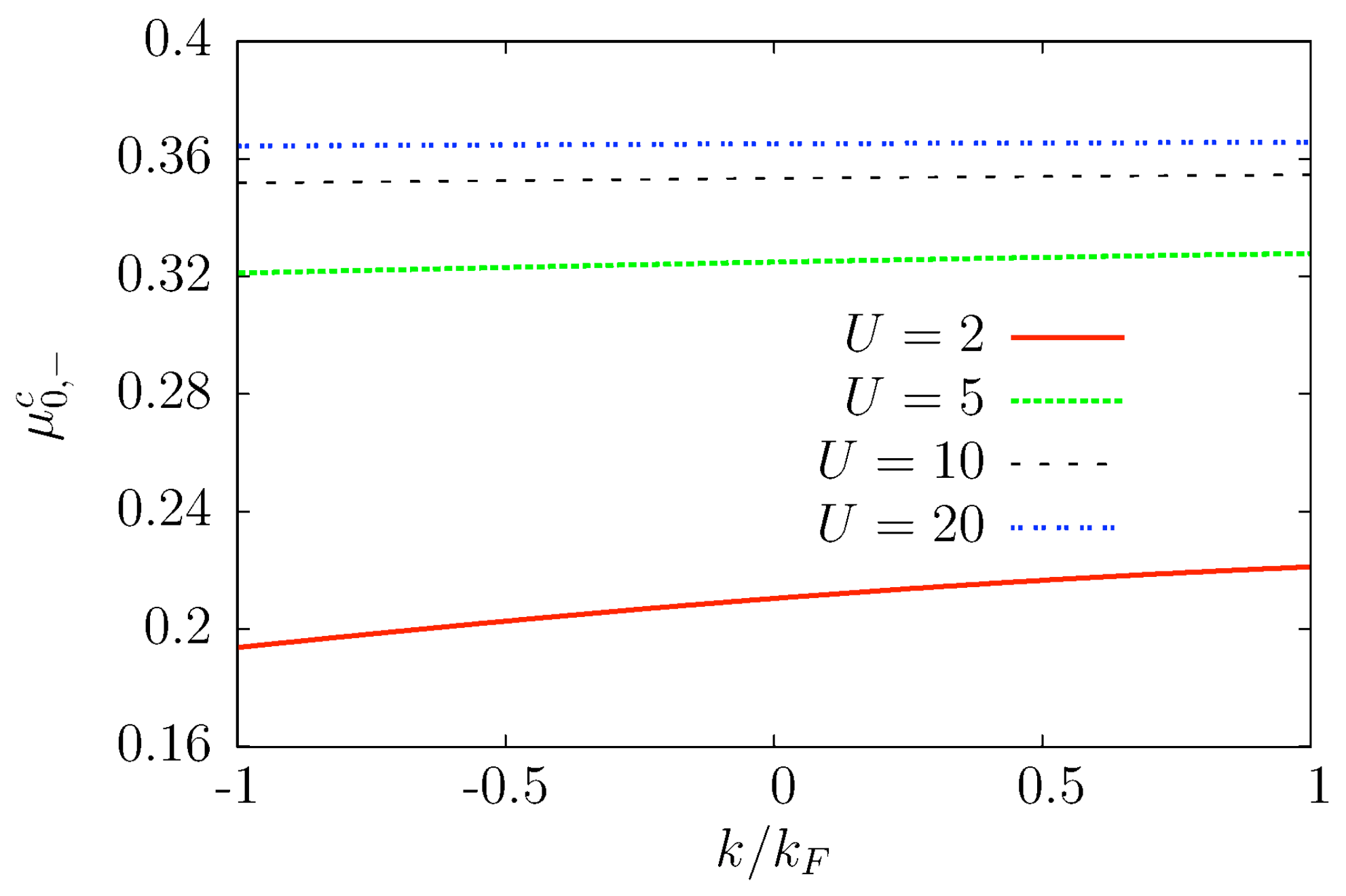}
\end{center}
\caption{(Color online)  
Exponent $\mu_{0,-}^c$ for $\rho=1/6$ as a function of momentum as
derived from the mobile impurity model approach for various fillings
and values of $U$.} 
\label{fig:rhop33}
\end{figure}
%
%
The exponent $\mu^c_{0,-}$ can then be calculated in
the framework of a mobile impurity model using input from the Bethe
Ansatz solution. Some details of this calculation are given in Appendix
\ref{app:A}. The result for the threshold exponent is
\be 
\mu_{0,-}^c = \frac{1}{2} - K_c \left( \frac{1}{2} - 2 D_c^\imp
\right) ^2 - \frac{\left( N_c^\imp \right)^2}{4K_c}\ ,
\label{eq:finalExp}
\ee
where $K_c$ is the Luttinger liquid parameter (cf (119) and (103)
of [\onlinecite{FHLE}]),
\bea
N_c^\mathrm{imp}&=&\int_{-Q}^Q\d k\ \rho_{c,1}(k)\ ,\nn
2D_c^\mathrm{imp}&=&\Phi(k^h)+
 \int_Q^\pi\d k\ \left[ \rho_{c,1}(-k) -
  \rho_{c,1}(k) \right]\nn
&-&\int_{-Q}^Q \d k\ \rho_{c,1}(k)\ \Phi(k), \nn
\Phi(k)&=&\frac{i}{\pi} \ln \left[ \frac{ \Gamma\left( \frac{1}{2} +
    i\frac{\sin k}{4u} \right)\Gamma\left( 1 - i\frac{\sin k}{4u}
    \right)}{\Gamma\left( \frac{1}{2} - i\frac{\sin k}{4u}
    \right)\Gamma\left( 1+ i\frac{\sin k}{4u} \right)} \right],
\eea
and the function $\rho_{c,1}$ is the solution of the integral equation
\bea 
\rho_{c,1}(k) =& -& \cos k\ R(\sin k - \sin k^h) \nn
&+& \cos k \int_{-Q}^Q \d
k' R(\sin k - \sin k') \rho_{c,1}(k').
\eea
Here $u=U/4$ and
\be
R(x)=\int_{-\infty}^\infty \frac{d\omega}{2\pi}\frac{e^{i\omega x}}
{1+\exp(2u|\omega|)}.
\label{rofx}
\ee

In Figs.~\ref{fig:rhop25} and \ref{fig:rhop33} we plot the value of the
exponent $\mu_{0,-}^c$ as a function of momentum for several values of
interaction strength $U$ and band fillings $1/4$ and $1/6$ respectively.
We note that our results are again in accord with those of
Refs.~[\onlinecite{carmelo}].

\section{Mobile impurity model and real-time dynamics} \label{sec:impurity}

We have seen above that for a given momentum \mbox{$0<|k|<k_F$} the
single-particle spectral function exhibits threshold
singularities at frequencies
\bea
\omega_1&=&E_{\rm hs}(Q,\Lambda^h)\ ,\quad P_{\rm
  hs}(Q,\Lambda^h)=k,\nn
\omega_2&=&E_{\rm hs}(k^h,-\infty)\ ,\quad P_{\rm
  hs}(k^h,-\infty)=k,\nn
\omega_3&=&E_{\rm hs}(k^h,-\infty)\ ,\quad P_{\rm hs}(k^h,-\infty)=-k.
\eea
Assuming that the singular features in the spectral function give the
dominant behaviour of the retarded Green's function at late times, we
conclude that the latter should be (approximately) of the form
\be
G(t,k)\sim\sum_\alpha A_\alpha e^{i\omega_\alpha t+\phi_\alpha} t^{-\gamma_\alpha},
\label{eq:fitting}
\ee
where the threshold exponents $\gamma_\alpha$ are related $\gamma_\alpha=\mu^{c/s}_{0,-}+1$ to the exponents calculated with the mobile impurity method. All parameters are a function of momentum $k$.
Here $A_\alpha$ are complex amplitudes and $\phi_\alpha$ are real phases.~It is
currently not known how to determine them {\it a~priori}, see however
Refs.~[\onlinecite{amplitudes}]. 
Table \ref{tab:explicitVals} gives explicit values for the frequencies $\omega_\alpha$ and exponents $\mu_{\alpha}$,  to be compared with TEBD results.
%
%
\begin{table}[t!]
\begin{tabular}{|c|c|c|c|c|c|c|c|} \hline \hline
$\rho$ & $U$ & $k$     & $-\omega_1$ & $-\omega_2$& $-\omega_3$ & $\mu_{0,-}^c$ & $\mu_{0,-}^s$ \\ \hline
1/4    &  5  &  0        & 0.387  & 1.15 & 1.15     & 0.373 & 0.855\\
1/4    &  5  & $\pi/8$ & 0.277 & 0.661   & 1.46  & 0.392 & 0.782\\
1/4    &  10 & 0       &  0.245  & 0.245   & 1.27  & 0.378 & 0.732\\
1/4    &  10 & $\pi/8$ & 0.178 & 0.713   & 1.62  & 0.388 & 0.660\\
1/6    &  5  &  0      & 0.165  &  0.564  &  0.564  & 0.325 & 0.851\\
1/6    &  5  & $\pi/8$ &  0.0631 & 0.173 & 0.745 & 0.237 & 0.632\\
1/6    &  10 & 0       &    0.101 &  0.644  & 0.644 & 0.353 & 0.710\\
1/6    &  10 & $\pi/8$ & 0.0384 & 0.195 & 0.856 & 0.354 & 0.545\\  \hline \hline
\end{tabular}
\caption{Values for the parameters of the power-law decay function~(\ref{eq:fitting}), calculated from the Bethe Ansatz and mobility impurity model.}
 \label{tab:explicitVals}
\end{table}
%
%
\begin{figure}[b!]
  \centering
  \includegraphics[width=\columnwidth]{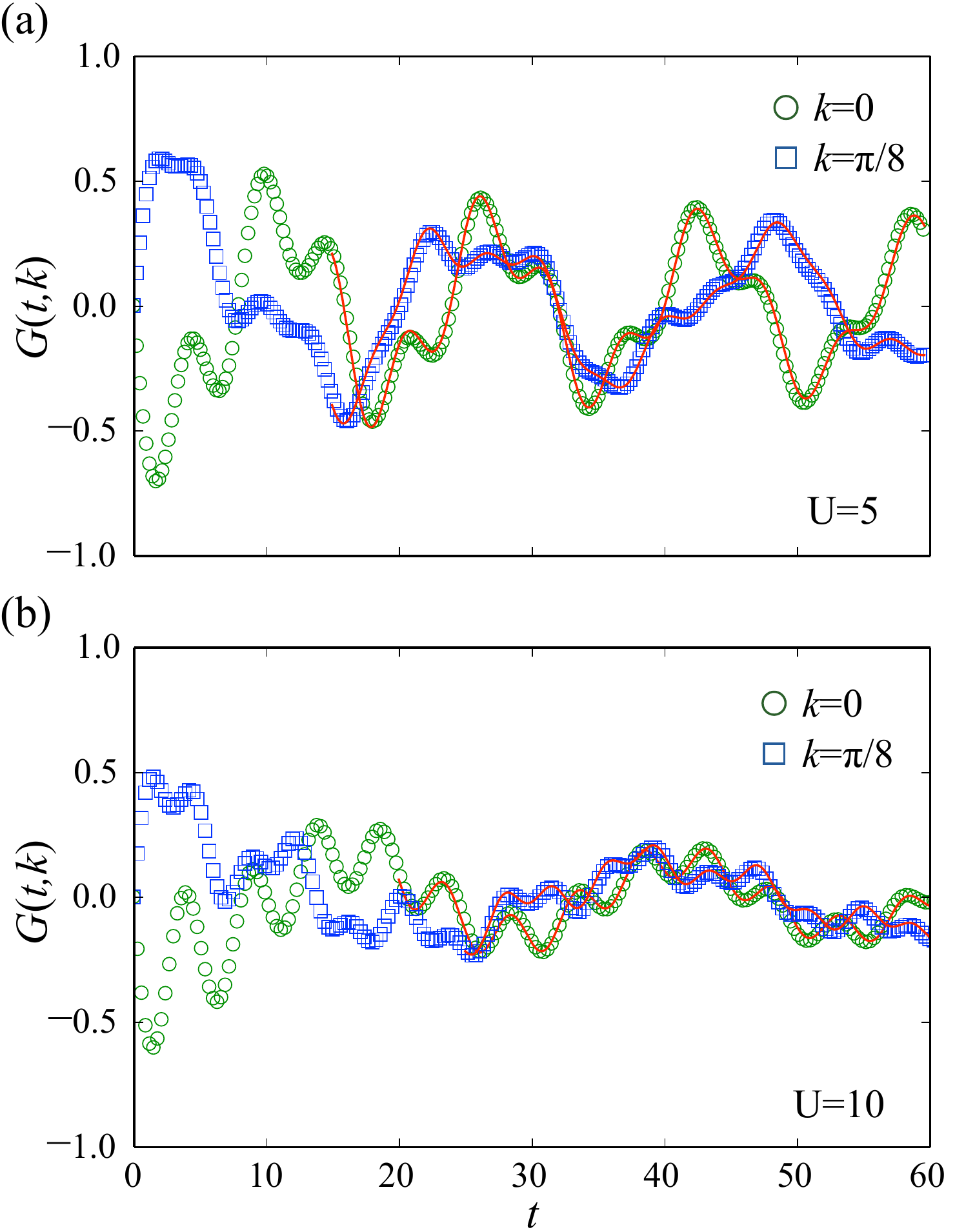}
  \caption{
(Color online)
Time decay of the imaginary part of the Green's function $G(t,k)$ for filling $\rho=1/4$.
Symbols are numerical TEBD data and red lines are fits to the power-law decay of the form of~(\ref{eq:fitting}).}
  \label{fig:fitn14} 
\end{figure}
%
\section{Comparison with numerical results}
\label{section:comparison}
%
%
%
%
\begin{figure}[b!]
  \centering
  \includegraphics[width=\columnwidth]{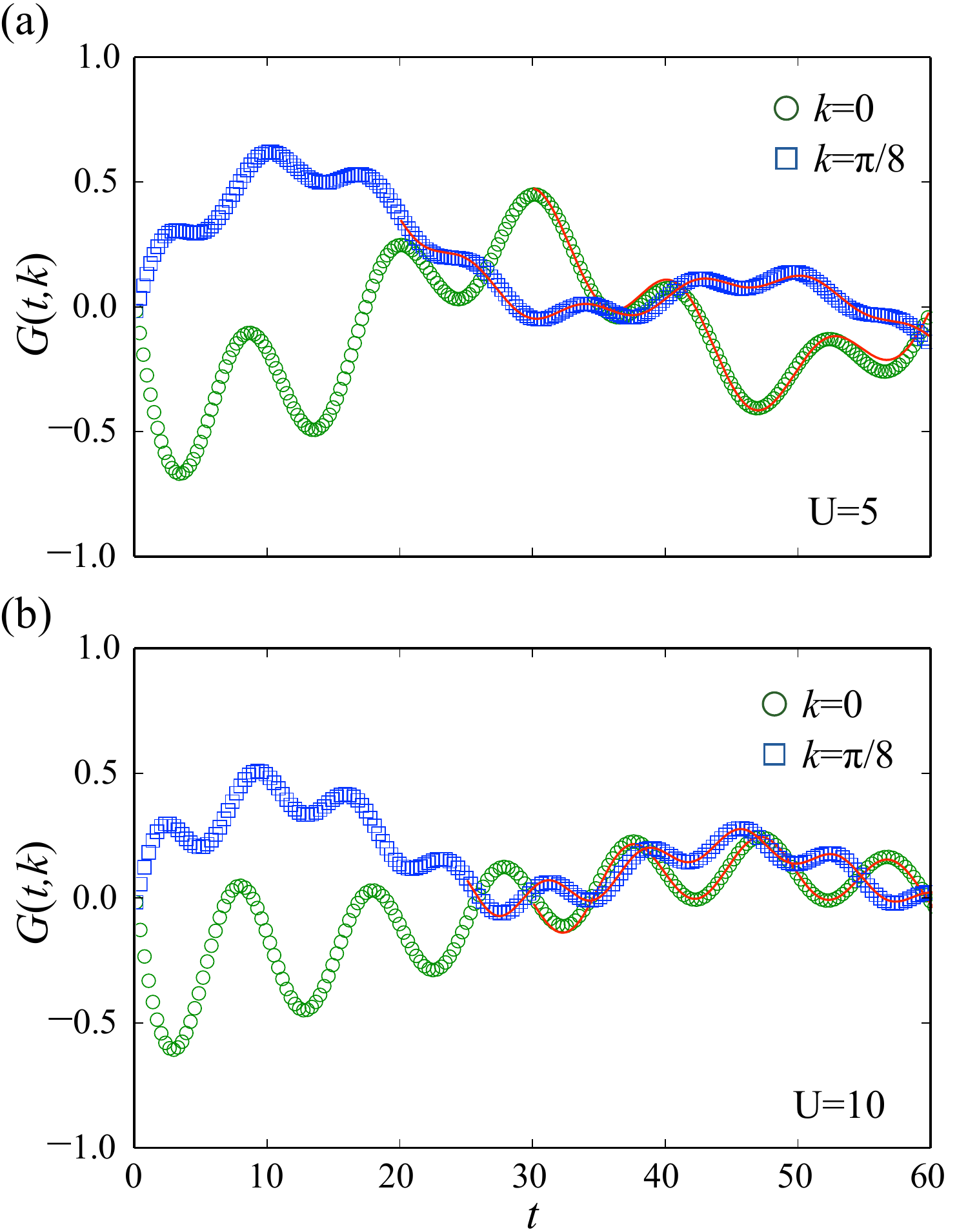}
  \caption{(Color online)
Time decay of the imaginary part of the Green's function $G(t,k)$ for filling $\rho=1/6$.
Symbols are numerical TEBD data and red lines are fits to the power-law decay of the form of~(\ref{eq:fitting}).}
  \label{fig:fitn16} 
\end{figure}

The formula~(\ref{eq:fitting}) for 
 $G(t,k)$  has  too many free parameters to reliably fit the numerical data from TEBD simulations, obtained through a Fourier transform of the (hole) Green's function~(\ref{eq:GFs}).
Therefore, we fix the threshold frequencies $\omega_\alpha$ and the  exponents $\gamma_\alpha$ to the values calculated with the mobile impurity approach, leaving the momentum-dependent amplitudes and phases as the only free parameters. 
We fit the imaginary part of $G(t,k)$ to the ansatz but the
procedure holds equally well with the real part. The data used
for the fitting procedure are not  extended in time with linear
prediction. 
The time evolution of the Green's function is illustrated for $\rho=1/4$  in Fig.~\ref{fig:fitn14} for $U=5$ and $U=10$, both at $k=0$ and $k=\pi/8$.
The initial time of the fit is adjusted in each case in order to avoid non-universal behaviour at short times.  For later times, the decay of $G(t,k)$ is very well reproduced by the fitting ansatz (\ref{eq:fitting}). 
The good quality of the fits to the numerical data is the main result of our paper, which validates the mobile impurity approach. 
As the momenta approaches $k_F$, the quality of the fit worsens slightly. 
One can understand this, since the frequency at which the first singularity develops approaches $\omega\rightarrow0$, which is more difficult to capture with a TEBD approach, inherently limited to a given finite time.

The same behaviour is essentially observed for  $\rho=1/6$ in Fig.~\ref{fig:fitn16}.
The relaxation to a universal behavior is slower for $\rho=1/6$, which agrees with the poorer resolution in frequency space observed in Fig.~\ref{fig:Akw-n14}.
Table \ref{tab:fittingVals} gives explicit values for the obtained fitting paramaters $A_\alpha$ and $\phi_\alpha$ for the range of parameters studied here.
%
%
\begin{table}[h!]
\begin{tabular}{|c|c|c|c|c|c|c|c|c|} \hline \hline
$\rho$ & $U$ & $k$     & $A_1$ & $A_2$& $A_3$ & $\phi_1$ & $\phi_2$ & $\phi_3$ \\ \hline
1/4    &  5  &  0      & 0.54  & 0.50  & 0.57 & 4.16 & 4.13 & 4.6\\
1/4    &  5  & $\pi/8$ & 0.57 & 0.71  & 0.34  & 1.03 & 5.1 & 3.74\\
1/4    &  10  &  0      & 0.39  & 0.94  & 1.02 & 3.83 & 0.25 & 3.62\\
1/4    &  10 & $\pi/8$ & 0.45 & 0.54  & 0.36  & 0.85 & 2.01 & 1.09\\
1/6    &  5  &  0      & 0.68  & 4.48  & 1.83 & 6.02 & 3.64 & 4.25\\
1/6    &  5  & $\pi/8$ & 0.76 & 2.00  & 0.47  & 1.42 & 2.41 & 6.06\\
1/6    &  10  & 0 &  0.46   & 2.38  & 3.29 & 4.62 & 3.14 & 0.38\\
1/6    &  10 & $\pi/8$ & 1.55 & 0.64  & 0.31  & 1.66 & 2.69 & 1.34\\
 \hline \hline
\end{tabular}
\caption{Fiting parameters of the power-law decay function~(\ref{eq:fitting}) to the time-decay of the Green's function calculated numerically with TEBD.}
 \label{tab:fittingVals}
\end{table}

As the fitting approach nicely reproduces the long-time behavior, we can use it instead of linear prediction to extend the raw TEBD data to longer times.
In the combined TEBD+BA approach we extrapolate the real-time data by orders of magnitude using the ansatz (\ref{eq:fitting}) with parameters taken from the analytical approach and the fitting procedure outlined above. 
\begin{figure}[h!]
  \centering
  \includegraphics[width=\columnwidth]{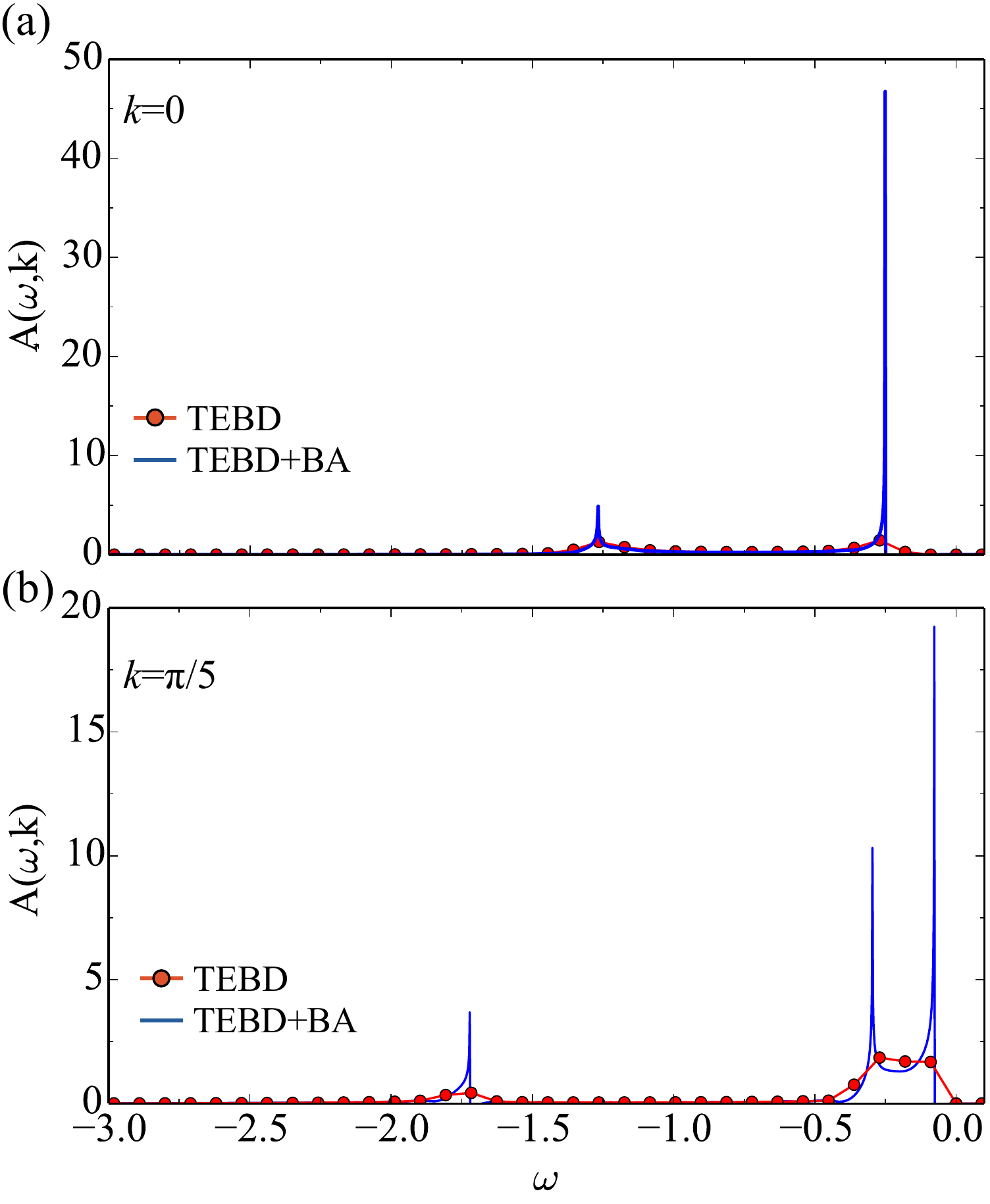}
  \caption{(Color online)
Comparison between the spectral function $A(\omega,k)$ at fixed momentum $k$ calculated with raw TEBD data (red) and TEBD extended in time with ansatz~(\ref{eq:fitting}) with BA and mobile impurity model parameters (blue). Results for quarter filling and $U=10$.}
  \label{fig:comparison} 
\end{figure}
Such a combined method dramatically increases the frequency resolution of the resulting spectral function with clearly defined singularity peaks, as show in Fig.~\ref{fig:comparison}, for the case of $\rho=1/4$ and $U=10$.
The spectral functions from both approaches have been normalized to obey the momentum-distribution function sum rule \mbox{$\int^0_{-\infty} A(\omega,k)=n(k)$}.

\section{Conclusions}
\label{sec:conclusions}
%
In this work we have studied the single particle Green's function
$G(t,k)$ of the one-dimensional repulsive Hubbard model in the
metallic regime. Using matrix product state techniques we 
computed $G(t,k)$ for a variety of band fillings and interaction
strengths for large systems ($L=300$) and times $0<t\alt 60$ (in units
of inverse hopping).
 We then employed mobile impurity models in tandem
with the Bethe Ansatz solution to obtain an expression for the late
time asymptotic behaviour of the Green's function. Using the unknown
coefficients in the resulting expression as fit parameters, we
obtained an excellent agreement with the numerical results at late
times.
 This strongly suggests that the mobile impurity results are not
only correct, but are of practical value. 
Moreover it removes concerns
based on the poor agreement between threshold singularity exponents
(in the frequency domain) obtained by mobile impurity models
\cite{FHLE} and earlier dynamical DMRG computations\cite{jeckel}.
Finally, we have shown that, combining the numerical results at short
and intermediate times with the asymptotic form dictated by the
mobile impurity model, it is possible to obtain results for the single
particle spectral function with unprecedented frequency resolution. 
We
expect this observation to be of practical use also in further cases.

\acknowledgments
This work was supported in part by the EPSRC under grants EP/I032487/1
(FHLE and TV) and EP/J014885/1 (FHLE and TV), and by the DFG under the grant SCHN 1169/2-1 (IS).

\appendix

\section{Calculation of the Holon Edge Exponent}
\label{app:A}
In this appendix we summarize the technical details underlying our
calculation of edge exponents in the framework of mobile impurity
models. In order to simplify the notation, we introduce the convolution $*$
\bea
K*f\Big|_k&=&\int_{-Q}^Q dq\ K(k,q)\ f(q)\ ,\nn
K(k,q)&=&\cos(k)\ R\big(\sin(k)-\sin(q)\big),
\label{K}
\eea
where the function $R(x)$ is given in \fr{rofx}. The action of the
transposed integral operator $K^T$ is defined by replacing $K(k,q)$ by
$K(q,k)$ on the right hand side of \fr{K}.
\subsection{Finite-size energy levels}
A key input in determining edge exponents is the finite-size
excitation spectrum in presence of a high-energy holon. The latter can
be obtained in a complete analogy to the calculation of the excitation
spectrum in the vicinity of the spinon edge in Ref.~[\onlinecite{FHLE}].
The result of this calculation~is
\begin{widetext}
\bea 
E(\Delta N_c, \Delta N_s, D_c, D_s) &=&
Le_{\mathrm{GS}} - \varepsilon_c(k^h) - \frac{\pi}{6L} (v_s + v_c) -
\frac{1}{L} \varepsilon_c'(k^h) \delta k^h \nn
&+&
\frac{2\pi v_c}{L} \left\{ \frac{ (\Delta N_c - N_c^\mathrm{imp})^2}{4\xi^2} + \xi^2\left( D_c - D_c^\mathrm{imp} + \frac{D_s}{2} - \frac{D_s^\mathrm{imp}}{2} \right)^2 \right\} \nn
&+&\frac{2\pi v_s}{L} \left\{ \frac{1}{2} \left( \Delta N_s -
\frac{\Delta N_c}{2} - N_s^\mathrm{imp} + \frac{N_c^\mathrm{imp}}{2}
\right)^2 + \frac{(D_s - D_s^\mathrm{imp})^2}{2} \right\}. 
\label{FSenergy}
\eea
\end{widetext}
The various quantities entering \fr{FSenergy} are as follows

\noindent (1)
The dressed energy for holons $\varepsilon_c(k)$ is a solution to the
integral equation 
\bea
\eps_c(k)&=&-2\cos(k)-\mu-2u+K^T*\eps_c\Big|_k.
\label{epsc}
\eea
The integration boundary $Q$ is fixed by the requirement $\eps_c(\pm
Q)=0$.

\noindent(2)
$\eps_c'(k)$ is the derivative of $\eps_c(k)$ with respect to $k$;

\noindent(3)
$e_{\rm GS}$ is the ground state energy per site
\be
e_{\rm GS}=\int_{-Q}^Q\frac{dk}{2\pi}\ \eps_c(k)+u\ ;
\ee

\noindent(4) The spin and charge velocities $v_{s,c}$ are obtained as
\bea
v_s&=&\frac{\eps_s'(\infty)}{2\pi \rho_s(\infty)}\ ,\qquad
v_c=\frac{\eps_c'(Q)}{2\pi \rho_c(Q)}\ ,\nn
\eps_s(\l)&=&\int_{-Q}^Qdk\cos(k)\ s(\l-\sin(k))\ \eps_c(k)\ ,\nn
\rho_c(k)&=&\frac{1}{2\pi}+K*\rho_c\Big|_k\ ,\nn
\rho_s(\l)&=&\int_{-Q}^Qdk\ s(\l-\sin(k))\ \rho_c(k)\ ;
\label{velocities}
\eea
where $s(x)=\big[4u\cosh\big(\frac{\pi x}{2u}\big)\big]^{-1}$.

\noindent(5)
The dressed charge $\xi=\xi(Q)$ is obtained from the solution of the
integral equation 
\be
\xi(k)=1+K^T*\xi\Big|_k\ .
\label{xiB0}
\ee
\noindent(6) $\Delta N_{c,s}$ and $D_{c,s}$ are quantum numbers describing
the excitation under consideration. If we only have a high-energy
holon with rapidity $k^h$ and a low-energy spinon sitting at its Fermi
point $\Lambda^h=-\infty$, then
\be
\Delta N_c=0\ ,\ \Delta N_s=-1\ ,\ D_s=-D_c=\frac{1}{2}.
\label{quantumnumbers}
\ee
The identification \fr{quantumnumbers} follows from the definition of $\Delta
N_{c,s}$ and $D_{c,s}$ in terms of the (half-odd) integers
characterizing a given solution of the Bethe Ansatz equation, cf
Ref.~[\onlinecite{FHLE}];

\noindent(7)
The quantities $N_{c,s}^{\rm imp}$ and $D_{c,s}^{\rm imp}$ are
given by
\bea
N_c^\mathrm{imp}&=&\int_{-Q}^Q\d k\ \rho_{c,1}(k)\ ,\nn
N_s^\mathrm{imp}&=&\int_{-\infty}^\infty \d
\Lambda\ \rho_{s,1}(\Lambda)=\frac{1}{2}\left(N^{\rm imp}_c-1\right)\ ,
\eea
\begin{widetext}
\bea
2D_c^\mathrm{imp}&=& \int_Q^\pi\d k\ \left[ \rho_{c,1}(-k) - \rho_{c,1}(k) \right] +
		\frac{i}{\pi} \ln \left[ \frac{ \Gamma\left( \frac{1}{2} + i\frac{\sin k^h}{4u} \right)\Gamma\left( 1 - i\frac{\sin k^h}{4u} \right)}{\Gamma\left( \frac{1}{2} - i\frac{\sin k^h}{4u} \right)\Gamma\left( 1+ i\frac{\sin k^h}{4u} \right)} \right]\nn 
&&-\frac{i}{\pi} \int_{-Q}^Q \d k\ \rho_{c,1}(k)\ \ln \left[ \frac{
    \Gamma\left( \frac{1}{2} + i\frac{\sin k}{4u} \right)\Gamma\left(
    1 - i\frac{\sin k}{4u} \right)}{\Gamma\left( \frac{1}{2} -
    i\frac{\sin k}{4u} \right)\Gamma\left( 1+ i\frac{\sin k}{4u}
    \right)} \right],\nn
D_s^{\rm imp}&=&0;
\eea
\end{widetext}
\noindent(8)
The ``order 1'' part $k^h$ of the holon rapidity is determined
by the requirement
\bea
z_c(k^h)&=&k^h + \int_{-\infty}^\infty \d
\Lambda\ \rho_s(\Lambda)\ \theta\left( \frac{\sin k^h - \Lambda}{u} 
\right)\nn
&=&\frac{2\pi I^h}{L}\ ,
\eea
where $I^h$ is a (half-odd) integer number characterizing the momentum
of the holon;

\noindent(9) The parameter $\delta k^h$ describes the ``order $1/L$'' part
of holon rapidity in the finite volume, and for zero magnetic field is
given by
\begin{widetext}
\bea 
\rho_{c,0}(k^h) \delta k^h&=&- \int_{-\infty}^\infty\frac{\d\Lambda}{2\pi}\
\rho_{s,1}(\Lambda)\ \theta\Big( \frac{ \sin k^h - \Lambda}{u}
\Big)
- \int_{-\infty}^\infty \frac{\d\Lambda}{2\pi}\sum_{\sigma=\pm} r_{sc}^{(\sigma)}(\Lambda)\ \theta\Big(\frac{\sin k^h
- \Lambda}{u}\Big)
(Q_\sigma-\sigma Q) + \frac{1}{2\sqrt{2}},
\eea
\end{widetext}
where the quantities $r_{\alpha c}^{(\sigma)}(z)$ satisfy
\bea
r_{cc}^{(\sigma)}(k) &=& K*
r_{cc}^{(\sigma)}\Big|_k+\sigma K(k,\sigma Q)\ ,\nn
r_{sc}^{(\sigma)}(\Lambda) &=& 
\int_{-Q}^Q \d k\, s(\Lambda - \sin k)\, r_{cc}^{(\sigma)}(k)\nn
&&+\sigma s(\Lambda - \sigma \sin Q).
\eea

\subsection{Impurity model and field theory}

The appropriate mobile impurity model for describing the holon edge
is\cite{SIG} $H_c+H_s+H_{\rm int}+H_d$, where
\be
H_\a=\frac{v_\a}{2\pi}\int dx\left[
\frac{1}{K_\a}\Big(\frac{\partial \Phi_\a}{\partial x}\Big)^2
+K_\a\Big(\frac{\partial \Theta_\a}{\partial x}\Big)^2\right]\ ,
\ee
\bea
H_{\rm int}&=&\int dx\ \left[\frac{V_R-V_L}{2\pi}\partial_x\Theta_c
-\frac{V_R+V_L}{2\pi}\partial_x\Phi_c \right]dd^\dagger,\nn
H_d&=&\int dx\ d^\dagger(x)\left[\varepsilon_c(P)-iu\partial_x\right]d(x)\ ,\nn
\label{HMI}
\eea
Here the Bose fields $\Phi_\a$ and the dual fields
$\Theta_\a$ fulfill the commutation relations 
$\left[\Phi_\a(x),\frac{\partial\Theta_\b(y)}{\partial
    y}\right]=i\pi\delta_{\a\b}\delta(x-y)$, $d(x)$ and
$d^\dagger(x)$ are annihilation and creation operators of the mobile
impurity, which carries momentum $P$ and travels at velocity $u$.
The parameters $V_{R,L}$ characterize the interaction of the impurity
with the low energy charge degrees of freedom. The parameters of
$H_{c,s}$ and $H_d$ in \fr{HMI} are readily identified with quantities
obtained from the Bethe Ansatz. The spin and charge velocities
$v_{s,c}$ are given by \fr{velocities} and the Luttinger parameters are
\be
K_s=1\ ,\quad K_c=\frac{\xi^2}{2},
\ee
where $\xi$ is given by \fr{xiB0}. The velocity of the impurity is
expressed in terms of the solutions to the integral equations
\fr{epsc}, \fr{velocities} as 
\be
u=\frac{\eps'_c(k^h)}{2\pi \rho_c(k^h)}.
\ee
The position $k^h$ of the hole is fixed by the requirement
\be
P_{hs}(k^h)=P.
\ee
The parameters $V_{R,L}$ entering the expression for $H_{\rm int}$ are
determined as follows. Following Ref.~[\onlinecite{SIG}] we remove the
interaction term $H_{\rm int}$ through the unitary transformation
$U^\dagger H U$
\bea
U^\dagger=\exp \Bigg( -i\int dx\Big[ 
\sqrt{K_c}\frac{\Delta\delta_{+,c}-\Delta\delta_{-,c}}{2\pi}\Theta_c(x)\nn
-\frac{\Delta\delta_{+,c}+\Delta\delta_{-,c}}{2\pi\sqrt{K_c}}\Phi_c(x)
\Big]d(x)d^\dagger(x) \Bigg),\nn
\label{udagger}
\eea
where
\be
(V_L\mp V_R)K_c^{\mp\frac{1}{2}}=(v_c+u)\Delta\delta_{-,c}\pm(v_c-u)\Delta\delta_{+,c}.
\ee
In the resulting Hamiltonian the impurity no longer interacts
explicitly with the charge part of Luttinger liquid, which in the
transformed basis takes the form
\bea
U^\dagger H_cU&=&\frac{v_c}{2\pi}\int dx\left[
\frac{1}{K_c}\left(\frac{\partial \widehat{\Phi}_c}{\partial x}\right)^2
+K_c\left(\frac{\partial \widehat{\Theta}_c}{\partial x}\right)^2\right]\ .\nn
\label{Hc}
\eea
The main effect of the unitary transformation is to change the
boundary conditions of the charge boson. In particular one has
\bea
\partial_x\widehat{\Phi}_c&=&U^\dagger\partial_x\Phi_cU
=\partial_x\Phi_c+\frac{\sqrt{K_c}}{2}
\Big(\Delta\delta_{+,c}-\Delta\delta_{-,c}\Big)dd^\dagger\ ,\nn
\partial_x\widehat{\Theta}_c&=&
U^\dagger\partial_x\Theta_cU=\partial_x\Theta_c-\frac{1}{2\sqrt{K_c}}
\Big(\Delta\delta_{+,c}+\Delta\delta_{-,c}\Big)dd^\dagger .\nn
\label{shifts}
\eea
Equations \fr{shifts} imply that the finite-size spectrum of \fr{Hc} in
presence of a high-energy holon impurity can be cast in the form
\begin{widetext}
\bea
\Delta E&=&\frac{2\pi v_c}{L}\left[
\frac{
\big(m_c+\bar{m}_c+\sqrt{2K_c}\frac{\Delta\delta_{c,+}-\Delta\delta_{c,-}}{2\pi}\big)^2}
{8K_c}
+\frac{K_c}{8}\big(m_c-\bar{m}_c
-\sqrt{\frac{2}{K_c}}\frac{\Delta\delta_{c,+}+\Delta\delta_{c,-}}{2\pi}
\big)^2
+\sum_{n>0}n\left[M_{n,c}^++M_{n,c}^-\right]
\right]\nn
&+&\frac{2\pi v_s}{L}\left[
\left(\frac{m_s}{2}\right)^2+\left(\frac{\bar{m}_s}{2}\right)^2
+\sum_{n>0}n\left[M_{n,s}^++M_{n,s}^-\right]\right],
\label{ELL}
\eea
\end{widetext}
see e.g. Ref.~[\onlinecite{CEL}], where 
\bea
m_\alpha+\bar{m}_\alpha&=&
-\frac{\sqrt{2}}{\pi}\int dx\ \langle\partial_x\Phi_\alpha\rangle\ ,\nn
m_\alpha-\bar{m}_\alpha&=&\frac{\sqrt{2}}{\pi}\int
dx\ \langle\partial_x\Theta_\alpha\rangle , \quad \alpha=c,s.
\eea
For the holon edge threshold we have
\be
m_c=\bar{m}_c=0\ ,\quad
m_s=-1\ ,\quad \bar{m}_s=0.
\ee
Comparing the resulting energies to the Bethe Ansatz form
\fr{FSenergy} we conclude that
\bea
N_c^{\rm imp}&=&
-\sqrt{2K_c}\frac{\Delta\delta_{c,+}-\Delta\delta_{c,-}}{2\pi}\ ,\nn
2D_c^{\rm  imp}&=&-\frac{1}{2}+
\frac{1}{\sqrt{2K_c}}\frac{\Delta\delta_{c,+}+\Delta\delta_{c,-}}{2\pi}\ . 
\eea

\subsection{Holon Edge Exponent}
Given the phase shifts $\Delta\delta_{c,\pm}$ the holon edge exponent
can be obtained following Ref.~[\onlinecite{SIG}]. The result is
\be
\mu_{0,-}^c = \frac{1}{2} - K_c \left( \frac{1}{2} - 2 D_c^\imp
\right) ^2 - \frac{\left( N_c^\imp \right)^2}{4K_c}.
\label{eq:finalExp2}
\ee
It is useful to consider particular limiting cases:
\begin{enumerate}
\item{}Infinite interaction limit $u\to\infty$
\be 
\lim_{u\to\infty} \mu_{0,-}^c = \frac{1}{2} - \frac{K_c}{4} =
\frac{3}{8}. 
\ee
As expected, this agrees with Ref. [\onlinecite{SIG}].
\item{}{$k\to k_F$ limit:}
here the result is
\be 
\lim_{k\to k_F}\mu_{0,-}^c= \frac{1}{2} - K_c \left( \frac{1}{2} -
\frac{1}{\sqrt{2K_c}} \right)^2 - \frac{ \left( 1 -
  \sqrt{2K_c}\right)^2}{4K_c} \ .
\ee
This again agrees with Ref.~[\onlinecite{SIG}], and is
\emph{different} from the Luttinger liquid result 
\be 
\mu_-^c = \frac{1}{2} - \frac{1}{8} \left( K_c + \frac{1}{K_c} - 2
\right). 
\ee
\end{enumerate}



\end{document}